\documentclass[%
superscriptaddress,
nofootinbib,
 amsmath,amssymb,
 aps,
onecolumn
]{revtex4-1}

\usepackage{color}
\usepackage{graphicx}
\usepackage{dcolumn}
\usepackage{bm}

\def\x{{\mathbf x}}%
\def\v{{\mathbf v}}%
\begin{document}


\title{Spacetime transformation acoustics}

\author{C. Garc\'{i}a-Meca}
 \affiliation{Nanophotonics Technology Center, Universitat Polit\`{e}cnica de Val\`{e}ncia, 46022 Valencia, Spain.}
 \email{Corresponding author. Email: cargarm2@ntc.upv.es}

\author{S. Carloni}%
\affiliation{ESA – Advanced Concepts Team, ESTEC, Keplerlaan 1, Postbus 299, 2200 AG Noordwijk, The Netherlands.}%

\author{C. Barcel\'{o}}
\affiliation{Instituto de Astrof\'{i}sica de Andaluc\'{i}a (CSIC), Glorieta de la Astronom\'{i}a, 18008 Granada, Spain.}

\author{G. Jannes}
\affiliation{Nanophotonics Technology Center, Universitat Polit\`{e}cnica de Val\`{e}ncia, 46022 Valencia, Spain.}%

\author{J. S\'{a}nchez-Dehesa}
\affiliation{Wave Phenomena Group, Universitat Polit\`{e}cnica de Val\`{e}ncia, 46022 Valencia, Spain.}%

\author{A. Mart\'{i}nez}
\affiliation{Nanophotonics Technology Center, Universitat Polit\`{e}cnica de Val\`{e}ncia, 46022 Valencia, Spain.}%

\date{\today}

\begin{abstract}
A recently proposed analogue transformation method has allowed the extension of transformation acoustics to general spacetime transformations.
We analyze here in detail the differences between this new analogue transformation acoustics (ATA) method and the standard one (STA).
We show explicitly that STA is not suitable for transformations that mix space and time.
ATA takes as starting point the acoustic equation for the velocity potential, instead of that for the pressure as in STA.
This velocity-potential equation by itself already allows for some transformations mixing space and time, but not all of them.
We explicitly obtain the entire set of transformations that leave its form invariant. It is for the rest of transformations that ATA shows its true potential, allowing for building a transformation acoustics method that enables the full range of spacetime transformations. We provide an example of an important transformation which cannot be achieved with STA. Using this transformation, we design and simulate an acoustic frequency converter via the ATA approach. Furthermore, in those cases in which one can apply both the STA and ATA approaches, we study the different transformational properties of the corresponding physical quantities.

\end{abstract}

\maketitle


\section{\label{sec:Intro}Introduction}

The success of transformation optics,\cite{LEO06-SCI,PEN06-SCI,SHA08-SCI,CHE10-NP} together with the availability of artificial materials with tailor-made properties,\cite{WEG10-PT,ZHA10-CSR} has led researchers to explore the possibility of applying similar techniques in other branches of physics. Outside of optics, acoustics is probably the field in which the greatest advance has been achieved. The form-invariance of the acoustic equations under spatial transformations is used to obtain the material parameters that deform acoustic space in the desired way. One of the most important applications of this technique is the cloaking of acoustic waves.\cite{CUM07-NJP, CHE07-APL, TOR07-NJP, NOR08-JASA, CHE10-JPD, POP11-PRL}

One possible problem in this process is that the transformations from physical to virtual space may result in metamaterials that cannot be
realized in practice. In order to overcome this problem, several authors~\cite{Urzhumov,Gokhale} have proposed to invert the process by first studying the range of realizable material parameters, and then deriving the appropriate transformations which guarantee the desirable effect, such as acoustic cloaking. Another problem, and the one that we will mainly address here, is related to the transformation process itself. Unlike electromagnetic theory, classical acoustics is based on non-relativistic equations that are non-invariant under transformations that mix space and time. As a consequence, the standard method for transformation acoustics cannot be applied to design devices based on this kind of transformation, contrarily to what has been done in optics.\cite{MCC11-JOPT,FRI12-NAT,CUM11-JOPT}

Recently, the construction of a general transformation acoustics formalism was tackled in a different way.\cite{GAR13-X,GAR12-ARIADNA} Instead of transforming directly the acoustic equations, the symmetries of an analogue abstract spacetime (described by relativistic equations) were exploited. In this method, each couple of solutions connected by a general coordinate transformation in the analogue spacetime can be mapped to acoustic space. In this way, it is possible to find the relation between the acoustic material parameters associated with each of these transformation-connected solutions. This method is referred to as analogue transformation acoustics (ATA) and revolves around the acoustic velocity potential wave equation and its formal equivalence with the relativistic equation that describes the evolution of a scalar field in a curved spacetime.\cite{VIS98-CQG,BAR11-LRR}

Since ATA and STA start from different initial equations (STA relies on pressure equations, whereas ATA starts from the velocity potential), it is worth studying the differences between the two methods. The first question that arises is whether it could be possible to construct an analogue transformation method based on the pressure wave equation, rather than the velocity potential formulation, and what its range of application would be. Second, it would be desirable to know if the pressure transforms in the same way in STA and ATA in those cases in which both methods can be used. Finally, we would like to explicitly obtain the set of transformations under which the acoustic equations are directly form-invariant in the original acoustic laboratory spacetime. In all the transformations that fall outside this set, the construction of the auxiliary relativistic analogue spacetime, and hence the use of ATA, is essential to achieve the desired transformation. All these questions are addressed in this work. In addition, to illustrate the potential of ATA, we analyze an example of a non-form-preserving transformation, namely, a space-dependent linear time dilation, which cannot be considered within STA. Using this transformation, we design and numerically test an acoustic frequency converter.

The paper is organized as follows. In section~\ref{s:Transformations} we outline the main limitation of the approach based on transforming directly the acoustic equations and present the set of transformations that do not preserve the form of the velocity potential equation (the detailed derivation can be found in appendix~\ref{s:Appendix}). In section~\ref{s:Review} we first review the ATA method. Then, we explicitly demonstrate that, although an analogue approach based on the pressure wave equation can in principle be constructed, it is not suitable for transformations that mix space and time. In section~\ref{s:Examples}, we design and analyze the above-mentioned acoustic frequency converter. The differences between STA and ATA are studied in depth in section~\ref{s:Pressure}. Finally, conclusions are drawn in section~\ref{s:Conclusion}.

\section{\label{s:Transformations}General spacetime transformations}

The various existing analyses in STA start from the following basic equation for the pressure perturbations $p$ of a (possibly anisotropic) fluid medium:\cite{HOOP}
\begin{align}
\label{pressure1}
\ddot p= B \nabla_i \left(\rho^{ij} \nabla_j p \right)~.
\end{align}
Here, $B$ is the bulk modulus and $\rho^{ij}$ the (in general, anisotropic) inverse matrix density of the background fluid. We will use latin spatial indices ($i,j$) and Greek spacetime indices ($\mu, \nu$, with $x^0=t$).
This is a Newtonian physics equation so that $\nabla$ represents the covariant derivative of the Newtonian flat 3-dimensional space. In generic spatial coordinates it will read
\begin{align}
\label{pressure}
\ddot p= B \frac{1}{\sqrt{\gamma}}\partial_i \left(\sqrt{\gamma} \rho^{ij} \partial_j p \right)~,
\end{align}
where $\gamma$ is the determinant of the three-dimensional spatial metric $\gamma_{ij}$ (with $\gamma^{ij}$ its inverse).
The success of STA relies on the form invariance of this equation under spatial coordinate transformations. It is easy to prove, however, that Eq.~\eqref{pressure} is not form invariant for more general (space-time mixing) transformations.

Another commonly used equation in acoustics is the one describing the evolution of the potential function $\phi_1$ for the velocity perturbation $\mathbf{v}_1$ defined as $\mathbf{v}_1=-\nabla\phi_1$~\cite{VIS98-CQG, BAR11-LRR}\footnote{Note that this definition does not impose any restriction on the vorticity of the background flow. In fact, even when the fluid is rotational, the present formalism can be maintained for sound waves satisfying $\omega\gg\omega_0$, with $\omega_0$ the rotation frequency of the background fluid and $\omega$ that of the acoustic perturbation. See the discussion in \cite{PER04-PD}.}:
\begin{eqnarray}
\label{wave_moving}
&-&\partial_t\left(\rho{c}^{-2}\left(\partial_t\phi_1 +\mathbf{v}\cdot\nabla\phi_1\right)\right) \\ \nonumber
&+&\nabla\cdot\left(\rho\nabla\phi_1-\rho c^{-2}\left(\partial_t\phi_1 + \mathbf{v}\cdot\nabla\phi_1 \right)\mathbf{v}\right)=0,
\end{eqnarray}
where $\mathbf{v}$ is the background velocity, $\rho$ the isotropic mass density and $c$ the local speed of sound ($B=\rho c^2$). This equation is in many cases equivalent to Eq.~\eqref{pressure}, but it is constructed using less stringent assumptions and naturally includes the velocity $\mathbf{v}$ of the background fluid. Therefore one could construct a transformation acoustics method based on this equation, which contains this additional degree of freedom $\mathbf{v}$.
In spite of this interesting feature, the use of  Eq.~\eqref{wave_moving} does not solve the problem of obtaining a transformation approach able to operate with spacetime transformations, since this equation is not invariant under general spacetime transformations either. Due to its complexity, it is not straightforward to see the exact set of transformations that do or do not preserve the form of Eq.~\eqref{wave_moving}. The first contribution of this work is the explicit derivation of these sets (see Appendix). By applying a generic spacetime coordinate transformation to Eq.~\eqref{wave_moving}, we can determine under which circumstances the form of this equation is preserved, without the appearance of new terms (which it would not be possible to interpret in terms of material parameters within the standard procedure). As a result, it is shown that form-invariance is satisfied whenever one of the following mutually exclusive sets of conditions holds (all the conditions of a set must hold simultaneously):
\begin{eqnarray}
\bullet \quad &&W_i=0, ~~ Z_{\bar i}=0.\\
\nonumber \\
\bullet \quad &&W_i\neq 0, ~~ Z_{\bar i}=0, \nabla_i W^i=0, \partial_t Z= 0 \mbox{ or } \partial_i Z= 0, \nonumber \\
&&\partial_t \sqrt{\gamma} =0  \mbox{ or } \partial_i\sqrt{\gamma}=0. \\
\nonumber \\
\bullet \quad&&W_i= 0, ~~ Z_{\bar i} \neq 0, ~~ \partial_t Z_{\bar i}=0,~~ \partial_t Z= 0, \nonumber \\
&&\partial_t \sqrt{\gamma} =0  \mbox{ or } \partial_i\sqrt{\gamma}=0. \\
\nonumber \\
\bullet \quad &&W_i \neq 0, ~~ Z_{\bar i} \neq 0, ~~ \nabla_i W^i=0,~~ \partial_t Z_{\bar i}= 0,~~ \partial_t Z= 0, \nonumber \\
&&\partial_t \sqrt{\gamma} =0  \mbox{ or } \partial_i\sqrt{\gamma}=0.
\end{eqnarray}
where we have defined the following elements
\begin{align}
W^i &= \frac{\partial \bar{x}^i}{\partial t}, \hspace{0.5cm}  Z_{\bar{i}}=\frac{\partial \bar{t}}{\partial x^i}, \hspace{0.5cm} Z = \frac{\partial \bar{t}}{\partial t}.
\end{align}
As can be seen, these conditions impose strong restrictions when it comes to mixing time with space. In fact, even a simple transformation such as a space-dependent linear time dilation does not belong to the kind of form-preserving mappings.

We can conclude that the standard transformational approach (based on a direct transformation of the equations) applied to either Eq.~\eqref{pressure} or Eq.~\eqref{wave_moving} does not allow us to work with most spacetime transformations. Therefore, another method is required.

\section{\label{s:Review}Analogue transformation acoustics}
In this Section, we will first review a different approach to the issue of transformation acoustics based on the formulation of an auxiliary relativistic theory, whose transformation properties will be the cornerstone of a new class of transformation approaches. The general method was presented in~\cite{GAR13-X}, here we restrict ourselves to its application to acoustics (see also the supplementary material to~\cite{GAR13-X}). This construction starts from the acoustic wave equation in terms of the velocity potential perturbation, Eq.~\eqref{wave_moving} above. Then, we will explicitly examine whether a similar analogue transformation method could have been obtained starting from the pressure wave equation Eq.~\eqref{pressure}, which has so far been the usual starting point in Transformation Acoustics.

\subsection{The Analogue Gravity equations}
In acoustics, the auxiliary model we need has been studied for some time and falls under the name of ``acoustic analogue gravity''. This analogue model tells us that  Eq.~\eqref{wave_moving} can be written as the relativistic equation of motion of a scalar field $\phi$ propagating in a (3+1)-dimensional pseudo--Riemannian manifold (the abstract spacetime), also called d'Alembert, Laplace-Beltrami or (massless) Klein-Gordon equation:\cite{VIS98-CQG, BAR11-LRR}
\begin{equation} \label{dAlembertian}
{1\over\sqrt{-g}} \partial_\mu \left( \sqrt{-g} \; g^{\mu\nu} \; \partial_\nu \phi \right)=0,
\end{equation}
where $g_{\mu\nu}$ is the 4-dimensional metric (with $g$ its determinant) of the abstract spacetime in which the field $\phi$ propagates. $g_{\mu\nu}$ is called sometimes the ``acoustic metric''.  Identifying $\phi$ and $\phi_1$ as analogue quantities, Eq.~\eqref{wave_moving} and Eq.~\eqref{dAlembertian} are mathematically identical provided that $g_{\mu\nu}$ is properly related to the acoustic parameters. Thus, the connection between Eq.~\eqref{wave_moving} and Eq.~\eqref{dAlembertian}  is provided by the elements of the metric.

The above formulation of acoustic analogue gravity has been developed with Eq.~\eqref{wave_moving} written in Cartesian coordinates, which typically are the most useful ``laboratory'' coordinates. However, it is easy to generalize this and demonstrate the formal equivalence of both equations when working in arbitrary spatial coordinates.
We start by noticing that Eq.~\eqref{wave_moving} for the perturbations of the velocity potential can be written in the form
\begin{equation}
\partial_t f^{00} \partial_t \phi + \partial_t f^{0i} \nabla_i \phi + \nabla_i f^{i0} \partial_t \phi + \nabla_i f^{ij} \nabla_j \phi=0~,
\end{equation}
with
\begin{eqnarray}
f^{\mu\nu} = {\rho \over c^2}
\left(
\begin{array}{ccc}
-1 & \vdots & -v^i \\
... & . & ................ \\
-v^i & \vdots & (c^2 \gamma^{ij}-v^iv^j)
\end{array}
\right)~.
\end{eqnarray}
Using the property
\begin{eqnarray}
\nabla_i V^i = {1 \over \sqrt{\gamma}} \partial_i (\sqrt{\gamma} V^i)~,
\end{eqnarray}
and realizing that $\sqrt{\gamma}$ is time-independent, we can write
\begin{equation}
{1 \over \sqrt{\gamma}}
\left(
\partial_t f_a^{00} \partial_t \phi + \partial_t f_a^{0i} \partial_i \phi + \partial_i f_a^{i0} \partial_t \phi + \partial_i f_a^{ij} \partial_j \phi
\right)=0~,
\end{equation}
where
\begin{eqnarray}
f_a^{\mu\nu} = {\rho \over c^2}\sqrt{\gamma}
\left(
\begin{array}{ccc}
-1 & \vdots & -v^i \\
... & . & ................ \\
-v^i & \vdots & (c^2 \gamma^{ij}-v^iv^j)
\end{array}
\right)~.
\end{eqnarray}
Dividing by $\rho^2/c$, this equation can be written as
\begin{eqnarray}\label{generalized_dalembertian}
{1 \over \sqrt{-g}}\partial_\mu \sqrt{-g}g^{\mu\nu} \partial_\nu \phi=0~,
\end{eqnarray}
provided that the metric is given by
\begin{eqnarray} \label{acoustic-metric}
g_{\mu\nu} = {\rho \over c}
\left(
\begin{array}{cccc}
-c^2+v^iv^j\gamma_{ij} &\vdots & -v^j\gamma_{ij}\\
................... & . & ........... \\
-v^j\gamma_{ij} &  \vdots & \gamma_{ij} & \\
\end{array}
\right)~,
\end{eqnarray}
such that the inverse metric and the metric determinant are
\begin{eqnarray} \label{inverse_4metric}
g^{\mu\nu} = \frac{1}{\rho c}
\left(
\begin{array}{ccc}
-1 & \vdots & -v^i \\
... & . & ................ \\
-v^i & \vdots & c^2\gamma^{ij}-v^iv^j
\end{array}
\right)~,
\end{eqnarray}
\begin{eqnarray}\label{metric_determinant}
\sqrt{-g}=\sqrt{\gamma} {\rho^2 \over c}~,
\end{eqnarray}
respectively.

A central issue in transformation physics is to have an expression valid (i.e., form-invariant) in a wide range of possible coordinates. Here we have seen that we can start from any arbitrary spatial coordinates and replace the original acoustic wave equation for the velocity potential \eqref{wave_moving} by a relativistic equation of the form \eqref{dAlembertian}. Beyond that, i.e., for arbitrary initial coordinates (obtained, e.g., from Cartesian coordinates through a transformation that mixes space and time), this formal equivalence is lost and one can no longer guarantee the equivalence between Eqs.~\eqref{wave_moving} and \eqref{dAlembertian}. However, it is crucial to realize that, once the mapping from Eq.~\eqref{wave_moving} to Eq.~\eqref{dAlembertian} has been performed (in an arbitrary {\it spatial} coordinate system), Eq.~\eqref{dAlembertian} will now remain form-invariant {\it under any arbitrary spacetime coordinate transformation}, precisely since \eqref{dAlembertian} is an
explicitly relativistic equation. This crucial observation leads us to define the following analogue method.

\subsection{The analogue method}
The ATA method is sketched in Fig.~\ref{fig:Figure_1} and consists of the following steps:
\begin{figure*}
\includegraphics{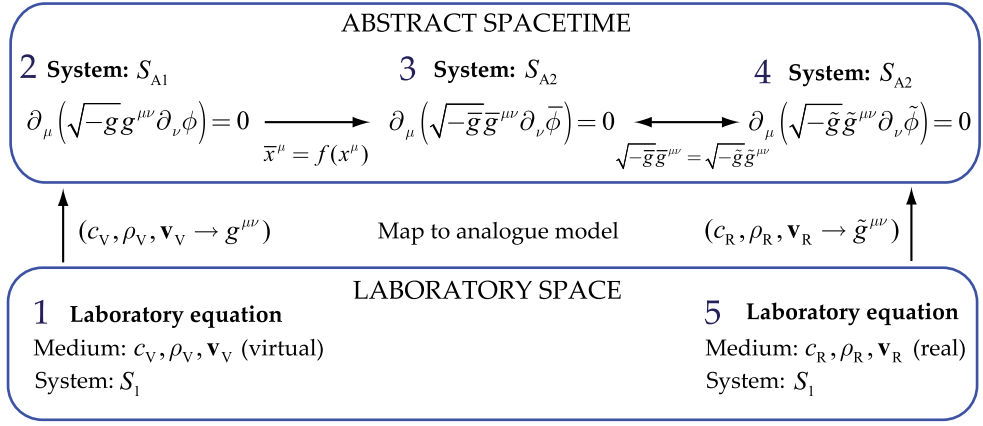}
\caption{\label{fig:Figure_1}Analogue Transformation Acoustics method.}
\end{figure*}
\begin{itemize}
\item  Start from a virtual medium of interest characterized by parameters $\rho_{\rm V}$, $c_{\rm V}$, and $\mathbf{v}_{\rm V}$
and express the (laboratory) acoustic equation in a coordinate system $S_1$ for which the relativistic analogy holds.
\item  Using Eq.~\eqref{inverse_4metric} particularized for the parameters of the virtual medium, derive its analogue model,
which is now a covariant equation in the abstract spacetime, expressed in a coordinate system $S_{\rm A1}$, with inverse metric
\begin{eqnarray}\label{g_V}
g^{\mu\nu} = \frac{1}{\rho_{\rm V} c_{\rm V}}
\left(
\begin{array}{ccc}
-1 & \vdots & -v_{\rm V}^i \\
... & . & ................ \\
-v_{\rm V}^i & \vdots & c_{\rm V}^2\tilde{\gamma}^{ij}-v_{\rm V}^iv_{\rm V}^j
\end{array}
\right)~.
\end{eqnarray}
\item  Perform the desired spacetime coordinate transformation $\bar{x}^{\mu}=f(x^{\mu})$ from system $S_{\rm A1}$ to another system $S_{\rm A2}$.
The new metric of step 3 is obtained, which follows from the one in step 2 by using standard tensorial transformation rules
\begin{equation} \label{metric_transformation}
\bar{g}^{\bar{\mu}\bar{\nu}}=\Lambda^{\bar{\mu}}_{\mu}\Lambda^{\bar{\nu}}_{\nu}g^{\mu\nu},
\end{equation}
where $\Lambda^{\bar{\mu}}_{\mu}=\partial{\bar{x}^{\bar{\mu}}}/\partial{x^{\mu}}$.
\item  Consider a second (real) medium $M_R$ characterized by parameters $\rho_{\rm R}$, $c_{\rm R}$, and $\mathbf{v}_{\rm R}$ (step 5) and derive its analogue model (step 4). Using Eq.~\eqref{inverse_4metric}, we know that the (inverse) metric associated to $M_{\rm R}$ will be
\begin{eqnarray}\label{g_R}
\tilde{g}^{\mu\nu} = \frac{1}{\rho_{\rm R} c_{\rm R}}
\left(
\begin{array}{ccc}
-1 & \vdots & -v_{\rm R}^i \\
... & . & ................ \\
-v_{\rm R}^i & \vdots & c_{\rm R}^2\tilde{\gamma}^{ij}-v_{\rm R}^iv_{\rm R}^j
\end{array}
\right)~.
\end{eqnarray}
\item  Finally, impose that the equations of steps 3 and 4 are equal (after relabeling $\bar{x}^{\mu}$ to $x^{\mu}$ in the expression for $\bar{g}^{\mu\nu}$), which implies that
\begin{equation} \label{ATA_condition}
\sqrt{-\bar{g}}\bar{g}^{\mu\nu}=\sqrt{-\tilde{g}}\tilde{g}^{\mu\nu}.
\end{equation}
From this equation we obtain the relation between the material parameters of the virtual and real media. The velocity potential in the medium $M_{\rm R}$ is the desired distorted version of that in $M_{\rm V}$.\cite{GAR13-X}
\end{itemize}
Using the ATA method, one can find the media $M_{\rm R}$ associated with a large set of transformations mixing space and time, which was not possible in STA.
For instance, all transformations mixing time with one spatial variable can be worked out. The only limitation comes from the fact that Eq.~\eqref{wave_moving} only considers isotropic fluids. A general transformation of coordinates has four arbitrary functions of spacetime, while the acoustic metric given by Eq.~\eqref{acoustic-metric} has only three independent functions $c(t,x)$, $\rho(t,x)$ and the background velocity potential $\phi_0(t,x)$, with $\mathbf{v}=-\nabla \phi_0(t,x)$. This constraint can be circumvented by incorporating anisotropy into Eq.~\eqref{wave_moving} through a homogenization technique.\cite{TOR09-PRB} For instance, the application of a two-scale homogenization method \cite{BEN78} to the velocity potential equation would result in a new equation for an effective acoustic system exhibiting an anisotropic mass density in the long-wavelength regime.

\subsection{ATA with the acoustic pressure wave equation}
In transformation acoustics there are two logically separate issues that should not be confused. One issue is whether one uses a pressure equation or a velocity potential equation. Another issue is whether one uses or not an intermediary abstract spacetime to perform the transformation. These two issues are combined in ATA as proposed so far only because the velocity potential equation is the one typically used in acoustics analogue gravity.     

Thus, the reader might wonder whether the ATA method would also work if Eq.~\eqref{pressure} was used instead of Eq.~\eqref{wave_moving}. As the second important contribution of this paper, let us show why this is not the case. As in the previous section, to construct such a method, we just need to obtain a connection between the original (not generally form-invariant) laboratory equation, in this case the pressure equation~\eqref{pressure}, and the relativistic equation~\eqref{dAlembertian}. Identifying $p$ and $\phi$ as analogue quantities, these two equations are mathematically identical when the metric $g_{\mu\nu}$ satisfies
\begin{align}
\label{metric3}
g^{\mu\nu}&=\left(\gamma\det{\left(\rho^{ij}\right)}B^{-1}\right)^{-\frac{1}{2}}
\left( \begin{array}{ccc}
-B^{-1} & \vdots & 0 \\
... & . & ................ \\
0 & \vdots & \rho^{ij}
\end{array} \right),\\
\label{determinant2}
g&=\det{\left(g_{\mu\nu}\right)}=-\gamma^2\det{\left(\rho^{ij}\right)}B^{-1}.
\end{align}
This result can be easily proven. Indeed, knowing that $g^{0i}=g^{i0}=0$, Eq.~\eqref{dAlembertian} becomes
\begin{align}
\label{dAlembertian3}
\ddot{p}=\frac{-1}{\sqrt{-g}g^{00}}\left(\sqrt{-g}g^{ij}p_{,j}\right)_{,i}.
\end{align}
Substituting the values of $g^{\mu\nu}$ and $g$ given by Eq.~\eqref{metric3} and \eqref{determinant2} into Eq.~\eqref{dAlembertian3}, we obtain
\begin{align}
\label{Wave_equation_spatial2}
\ddot{p}=\frac{\left(\gamma\det{\left(\rho^{ij}\right)}B\right)^{1/2}}{\gamma\left(\det{\left(\rho^{ij}\right)}B^{-1}\right)^{1/2}}\left(\frac{\gamma\left(\det{\left(\rho^{ij}\right)}B^{-1}\right)^{1/2}}{\left(\gamma\det{\left(\rho^{ij}\right)}B^{-1}\right)^{1/2}}\rho^{ij}p_{,j}\right)_{,i}.
\end{align}
After simplification,
\begin{align}
\label{Wave_equation_spatial3}
\ddot{p}=\frac{B}{\sqrt{\gamma}}\left(\sqrt{\gamma}\rho^{ij}p_{,j}\right)_{,i},
\end{align}
i.e. Eq.~\eqref{pressure}.
Therefore, we could use Eq.~\eqref{pressure} in laboratory space in Fig.~\ref{fig:Figure_1} and employ this analogy between Eqs.~\eqref{pressure} and ~\eqref{dAlembertian}. We would start from Eq.~\eqref{pressure} particularized for a virtual medium characterized by parameters $B_{\rm V}$ and $\rho_{\rm V}^{ij}$ (step 1) and obtain its analogue model (step 2) with an associated metric $g_{\mu\nu}$ given by Eq.~\eqref{metric3} particularized for the mentioned parameters. Then, apply the desired transformation, which results in the equation of step 3 with a transformed metric $\bar{g}_{\mu\nu}$.

However, for a transformation that mixes space and time, the required metric components $\bar{g}^{0i}$ and $\bar{g}^{i0}$ will be non-vanishing in general. But according to Eq.~\eqref{metric3}, the metric of step 4 should have the following form
\begin{align}
\label{metric4}
\tilde{g}^{\mu\nu}&=\left(\gamma\det{\left(\rho^{ij}\right)}B_{\rm R}^{-1}\right)^{-\frac{1}{2}}
\left( \begin{array}{ccc}
-B_{\rm R}^{-1} & \vdots & 0 \\
... & . & ................ \\
0 & \vdots & \rho_{\rm R}^{ij}
\end{array} \right).
\end{align}
Clearly, then, the condition expressed by Eq.~\eqref{ATA_condition} cannot be fulfilled, since $\tilde{g}^{0i} =\tilde{g}^{i0} = 0$. This is why a general spacetime transformation cannot be implemented with an acoustic system described by Eq.~\eqref{pressure}. On the contrary, the system described by Eq.~\eqref{wave_moving} has an equivalent metric with non-vanishing components $\tilde{g}^{0i}$ and $\tilde{g}^{i0}$. Note that these components would be zero if the background velocity were also zero, see Eqs.~\eqref{g_V} and \eqref{g_R}. Thus, allowing the background fluid to move is a crucial ingredient of the analogue transformation method. However, to our knowledge, there is no similar wave equation to Eq.~\eqref{wave_moving} for the pressure. This shows the importance of choosing the adequate variable to construct a complete transformation approach in this case.

The velocity potential equation by itself needs less assumptions for its
 validity (in particular, contrarily to the pressure wave equation~\cite{BER46-JASA}, it doesn't need a vanishing background pressure
 gradient). Also, it encompasses more configurations, first by explicitly
 incorporating background fluid flows, and second by allowing density
 gradients even with a homogeneous (non-space-dependent) equation of state.
 Moreover, historically it has been the natural starting point used in
 Analogue Gravity, while we have just seen that, although a similar
 analogue metric could be constructed starting from the pressure equation,
 this would not provide space-time mixing coefficients in the metric.
 For these reasons (see also Section~\ref{s:Pressure} and Ref.~\onlinecite{GAR13-X}),
 we use the name ATA explicitly for the combined use of the velocity
 potential equation and the analogue transformation philosophy.

%
\section{\label{s:Examples} Example: an acoustic frequency converter}
In this section, we will demonstrate that an acoustic frequency converter can be designed with ATA. This was not the case with STA, and therefore illustrates the strength of ATA (see also \cite{GAR13-X} for other examples of applications which can be designed with ATA but not with STA).

Let us consider the following transformation:
\begin{align}\label{converter_transformation}
\bar{t}&=t(1+ax),\\
\bar{x}^i&=x^i,
\end{align}
with $a$ having units of inverse length.
This is an interesting transformation that has been used in the context of
transformation optics to design frequency converters.~\cite{CUM11-JOPT}
Obviously, this transformation does not satisfy the form-invariance conditions. As
a consequence, the use of the ATA approach is indispensable in this case.
The relation between the parameters of real and virtual media for a general transformation \mbox{$\left(\bar{t}=f_1(x,t);\, \bar{x}=f_2(x,t)\right)$} mixing the $x$ and $t$ variables is given by~\cite{GAR13-X}
\begin{align}
v_{\rm R}^x&=\frac{\partial_t{f}_1\partial_t{f}_2-c_{\rm V}^2\partial_x{f}_1\partial_x{f}_2}  {(\partial_t{f}_1)^2-c_{\rm V}^2(\partial_x{f}_1)^2},\\
c_{\rm R}^2&=(v_{\rm R}^x)^2 + \frac{c_{\rm V}^2(\partial_x{f}_2)^2-(\partial_t{f}_2)^2} {(\partial_t{f}_1)^2-c_{\rm V}^2(\partial_x{f}_1)^2},\\
\rho_{\rm R}&=\rho_{\rm V}\frac{c_{\rm R}^2}{c_{\rm V}^2} \frac {(\partial_t{f}_1)^2-c_{\rm V}^2(\partial_x{f}_1)^2} {\partial_t{f}_1\partial_x{f}_2-\partial_x{f}_1\partial_t{f}_2}.
\end{align}
According to the previous equations, the transformation in Eq.~\eqref{converter_transformation} can be implemented by using the following parameters (after renaming $\bar{x}$, $\bar{t}$ to $x$, $t$):
\begin{align}\label{Frequency_converter_1}
v_{\rm R}^x&=\frac{-c_{\rm V}^2at(1+ax)}{(1+ax)^4-(c_{\rm V}at)^2}, \\
\frac{c_{\rm R}}{c_{\rm V}}&=\frac{\rho_{\rm R}}{\rho_{\rm V}}=\frac{\left(1+ax\right)^3}{\left(1+ax\right)^4-\left(c_{\rm V}at\right)^2}.
\label{Frequency_converter_2}
\end{align}
In a practical situation, this transformation is only applied in a certain region $0 < x <
L$. Taking $a=(m^{-1}-1)/L$, with $m$ a constant, we ensure that the transformation
is continuous at $x=0$. In this case, $\bar{t}=t/m$ at $x=L$. Continuity can be
guaranteed at $x=L$ by placing the medium that implements the transformation
$(\bar{t}=t/m;\, \bar{x}^i=x^i)$ at the device output ($x>L$) . It can easily be
shown that the properties of such a medium are
\begin{align}\label{Frequency_converter_output}
v_{\rm R}^x=0\, ; \qquad \frac{c_{\rm R}}{c_{\rm V}}=\frac{\rho_{\rm R}}{\rho_{\rm V}}=m.
\end{align}
Following a reasoning similar to that of Ref.~\onlinecite{CUM11-JOPT}, it is
possible to prove that, after going through the device, the acoustic signal
frequency is scaled by a factor of $m$. To that end, we assume that
the potential has the form of a plane wave with a space-dependent frequency
$\phi_1=\phi_c \exp\left(i\omega(x)t-ikx\right)$ (the problem is invariant in the $y$ and
$z$ directions). Substituting this \emph{ansatz} into the wave equation and neglecting $\partial_x^2 \omega$, we obtain the following relation
\begin{align}\label{freq_equation}
-\omega c^{-2}\left(\omega+v_x\Delta_x\right) + \Delta_x\left[\Delta_x-c^{-2}\left(w+v_x\Delta_x\right)\right] = 0,
\end{align}
with $\Delta_x=td\omega / dx-k$. Inserting Eqs.~\eqref{Frequency_converter_1}--\eqref{Frequency_converter_2} into Eq.~\eqref{freq_equation}, we arrive at
\begin{align}
\left[\frac{\omega\left(1+ax\right)}{c_{\rm V}}\right]^2=\left[\Delta_x+\frac{\omega at}{1+ax}\right]^2.
\end{align}
The solution to this equation is the following dispersion relation
\begin{align}
\omega(x)=\pm \frac{c_{\rm V}k}{1+ax},
\end{align}
from which it is clear that $\omega(x=L)=m\omega(x=0)$.

Note that, as in the optical case, a time-varying medium is required to achieve a frequency conversion, while  in static media only a wavelength change can be observed (as a consequence of the medium spatial variation), as for instance in the case analyzed in \cite{CHE10-OE}.

\begin{figure}[b]
\includegraphics{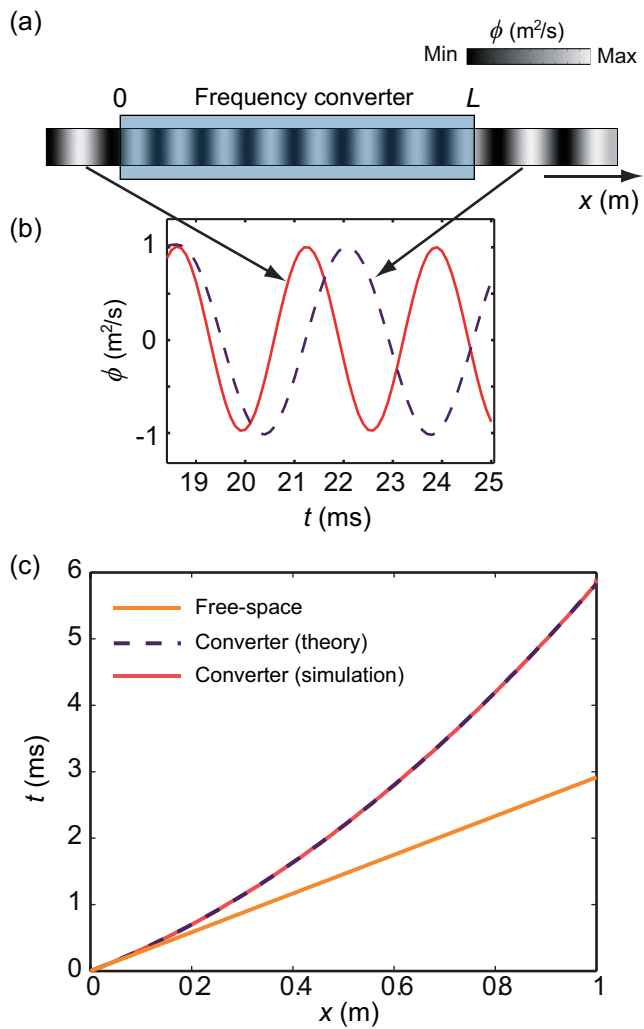}
\caption{\label{fig:Time_transformation} Acoustic frequency converter designed with ATA. (a) Velocity potential as a function of space at a given instant. (b) Time dependence of the velocity potential at two different positions to the left and right of the converter. (c) Trajectory followed by an acoustic ray inside a converter with $a=1$ m$^{-1}$.}
\end{figure}

In order to check the validity of the acoustic frequency converter, a particular case with $L = 5$ m and $m = 0.8$ was simulated numerically. The calculation was carried out by using the acoustic module of COMSOL Multiphysics, where the weak form of the aeroacoustic wave equation was modified to allow for time-varying density and speed of sound.
In the simulation, an acoustic wave impinges onto the frequency converter from the left. The space dependence of the velocity potential at a certain instant is depicted in Fig.~\ref{fig:Time_transformation}(a). It can be seen that, while the wavelength grows with $x$ inside the converter, it is the same at the input and output media. Since the speed of sound of the output medium is $m$ times that of the input medium, the output frequency must also be $m$ times the input one. This relation can also be obtained by comparing the time evolution of the velocity potential at two arbitrary positions to the left and right of the converter [see Fig.~\ref{fig:Time_transformation}(b)].
To further verify the functionality of the converter, we calculated the trajectory followed by an acoustic ray immersed in such a medium, starting at $x=0,t=0$. This was done by solving numerically Hamilton's equations (see Ref.~\onlinecite{GAR13-X} for further details). In this case we chose $a=1$ m$^{-1}$. The calculated time-position curve and the expected theoretical curve are depicted in Fig.~\ref{fig:Time_transformation}(c). Both are identical. For comparison purposes, the curve associated to the propagation of sound in the reference fluid is also shown.

To conclude this section, as a third contribution of this paper, we have demonstrated that an acoustic frequency-converter (the acoustic equivalent of the optical frequency-converter discussed in~\cite{CUM11-JOPT}), which was not possible in the standard formalism STA, can indeed be designed using ATA.

%
\section{\label{s:Pressure} Relation between STA and ATA}
%
In the case of pure spatial transformations, both ATA and STA are defined and can be used. However, since they are based on different wave equations, the results obtained with both approaches might not be completely equivalent. In this Section, as a fourth contribution of this paper, we will analyze the similarities and differences between both techniques. To avoid dealing with anisotropic materials [not supported by Eq.~\eqref{wave_moving}], we will focus on conformal transformations, which preserve isotropy. For simplicity, let us choose a coordinate system for laboratory space in which the spatial metric is
\begin{align}
\label{metric0}
\gamma_{ij}&=\Omega_{\rm R}\delta_{ij},
\end{align}
where $\Omega_{\rm R}$ is a function of the spatial coordinates. After a conformal transformation, the spatial part of the metric $\bar{g}_{\mu\nu}$ in step 3 of Fig.~\ref{fig:Figure_1} will be~\cite{GAR13-X}
\begin{align}
\label{metric1}
\bar{g}_{ij}=\frac{\rho_{\rm V}}{c_{\rm V}}\Omega_{\rm V}\delta_{ij},
\end{align}
where $\Omega_{\rm V}$ is also a function of the spatial coordinates. Note that $\Omega_{\rm V}\delta_{ij}$ represents either the spatial metric arising from a three-dimensional conformal (M\"{o}bius) transformation of flat space, or the metric of a curved --- but conformally flat --- space (for example, Maxwell's fisheye corresponds to the latter). Moreover, $g_{0i}=g_{i0}=0$. As discussed above, this particular form of the spacetime metric ensures that STA is defined and can be compared with ATA. In fact, this specific scenario was analyzed in detail in Ref.~\onlinecite{GAR13-X} and it was found that the virtual and real densities are related differently in each case:
\begin{align}
\label{Density_pressure2}
\rho_{\rm R_{STA}}&=\rho_{\rm V}\frac{\Omega_{\rm R}^{1/2}}{\Omega_{\rm V}^{1/2}},\\
\label{Density_potential2}
\rho_{\rm R_{ATA}}&=\rho_{\rm V}\frac{\Omega_{\rm V}^{1/2}}{\Omega_{\rm R}^{1/2}}.
\end{align}

Let us clarify the origin of this difference. For that, we must notice that Eqs.~\eqref{pressure} and~\eqref{wave_moving} are based on different assumptions. Specifically, to obtain density gradients with the pressure equation, one needs to assume a space-dependent equation of state.~\cite{BER46-JASA} This is not necessary with the velocity potential equation, where density gradients can come directly from gradients in the background pressure.~\cite{BAR11-LRR} As a consequence, both equations are not fully equivalent. This difference becomes explicit when we compare the pressure equation in the isotropic case with the velocity potential equation when all the background quantities are stationary (a dynamic background is not required for purely spatial transformations). In general spatial coordinates, these equations read
\begin{eqnarray}
\text{(STA)}~~~&&\partial_t^2 p =c^2 \rho~\partial_i\left(\rho^{-1}\sqrt{\gamma}\gamma^{ij}\partial_j p \right)~,
\\
\text{(ATA)}~~~&&\partial_t^2{\phi}=c^2 \rho^{-1} ~\partial_i\left(\rho \sqrt{\gamma} \gamma^{ij}\partial_j\phi_1 \right)~.
\end{eqnarray}
Using the following relation between the acoustic perturbations $p$ and $\phi_1$~\cite{BAR11-LRR}
\begin{eqnarray}
\label{Relation}
p=\rho \partial_t \phi_1,
\end{eqnarray}
we obtain
\begin{eqnarray}
\label{pressure_static}
\text{(STA)}~~~&&\partial_t^2 p =c^2 \rho~ \partial_i\left(\rho^{-1}\sqrt{\gamma}\gamma^{ij}\partial_j p \right)~,
\\
\text{(ATA)}~~~\label{potential_static}
&&\partial_t^2 p =c^2 ~\partial_i\left(\rho \sqrt{\gamma} \gamma^{ij}\partial_j (p/\rho) \right)~.
\end{eqnarray}
From here one can easily see that both equations coincide if and only if $\partial^2 \rho=0$. However, even when the condition $\partial^2 \rho=0$ is satisfied, the way in which one can use a density parameter to simulate a conformal factor in the spatial geometry is nevertheless different in both approaches. So, imagine for instance that $\rho$ is constant. Then, in the same system of Cartesian coordinates, the previous two equations are identical. However, in general, when trying to absorb a conformal transformation of coordinates into a physical parameter $\rho_R$, each approach leads to a different physical situation, and hence to a different transformed density, compare Eqs.~\eqref{Density_pressure2} and \eqref{Density_potential2}. Both are in principle workable but, depending on the particular situation and the available (meta)materials, one could be easier to implement than the other.

As a consequence of the different density distributions prescribed by STA and ATA, we expect that the acoustic pressure also transforms differently in each approach. Let us examine this difference. In STA we directly transform the equation for the pressure. Thus, since the pressure transforms as a scalar, the pressure perturbation in real space $\tilde{p}$ (Step 5) is related to the pressure perturbation in virtual space $p$ (Step 1) as
\begin{equation} \label{pressure_relation1}
\tilde{p}(t,x^i)=p(t,f^{-1}(x^i)),
\end{equation}
where $f$ is the spatial coordinate transformation performed to change from Step 2 to Step 3.
However, in ATA, we apply the transformation to the wave equation for the potential, so we have
\begin{equation}
\tilde{\phi}_1(t,x^i)=\phi_1(t,f^{-1}(x^i)),
\end{equation}
where $\phi$ and $\tilde{\phi}$ are the potentials in Steps 1 and 5, respectively. Now, using Eq.~\eqref{Relation},
we know that
\begin{equation}
p(t,x^i)=\rho_{\rm V} (x^i)\partial_t{\phi_1(t,x^i)},
\end{equation}
and therefore
\begin{eqnarray}
\label{pressure_relation}
\tilde{p}(t,x^i)&=&\rho_{\rm R}(x^i)\partial_t\tilde{\phi}_1(t,x^i) \nonumber \\
&=&\rho_{\rm V}(f^{-1}(x^i))\frac{\Omega_{\rm V}^{1/2}}{\Omega_{\rm R}^{1/2}}\partial_t\phi_1(t,f^{-1}(x^i)) \nonumber \\
&=&p(t,f^{-1}(x^i))\frac{\Omega_{\rm V}^{1/2}}{\Omega_{\rm R}^{1/2}}.
\end{eqnarray}
Comparing with the corresponding transformation in STA, Eq.~\eqref{pressure_relation1}, we can see that the pressure in ATA is further corrected by a
factor $\Omega_{\rm V}^{1/2}/\Omega_{\rm R}^{1/2}$.

The fact that the same coordinate transformation results in a different pressure transformation in STA and ATA is consistent with the different acoustic media prescribed by each method\footnote{The non-uniqueness of material parameters for acoustic transformations was studied, e.g., in \cite{NOR08-JASA} and emphasized in \cite{NOR11-WM}. There, this liberty came essentially from the possibility that both the bulk modulus and/or the density could become tensorial. Here, we have shown that, even when choosing the bulk modulus to remain scalar, different material parameters can still be obtained for a given transformation, depending on the formalism chosen.}. On the other hand, STA and ATA exhibit an important similarity in this case. In particular, both acoustic media will behave equally from a line-dragging perspective, or in other words, acoustic rays will follow the same trajectories in both cases. To see this, we look at the relation between the speed of sound of virtual and real media is~\cite{GAR13-X}
\begin{equation}
\label{Speed_potential}
c_{\rm R_{STA}}=c_{\rm R_{ATA}}=c_{\rm V}\frac{\Omega_{\rm R}^{1/2}}{\Omega_{\rm V}^{1/2}},
\end{equation}
i.e., this quantity transforms equally in both approaches. In order to understand this, we look again at Eqs.~\eqref{pressure_static} and~\eqref{potential_static}.
Working in Cartesian coordinates for simplicity and expanding the derivatives, the equations become
\begin{eqnarray}
&&\text{(STA)}~~~  \partial_t^2 p = c^2 \delta^{ij}\left[\partial_i\partial_j p + \rho \partial_i\left(\rho^{-1}\right)\partial_j p \right]~,\\
\label{potential_static_expanded}
&&\text{(ATA)}~~~ \partial_t^2 p = c^2 \delta^{ij} \nonumber \\
&& \times \left[\partial_i\partial_j p + \rho \partial_i\left(\rho^{-1}\right)\partial_j p + \partial_i\left(\rho \partial_j\rho^{-1}\right)p\right]~.
\end{eqnarray}
Thus, both equations are identical if the last term in brackets in Eq.~\eqref{potential_static_expanded} is negligible. This occurs at high frequencies, for which the spatial derivatives of $p$ are much larger than $p$ itself. The fact that in the geometrical approximation (high-frequency limit) both equations are equivalent is consistent with the fact that the speed of sound (the only relevant parameter in ray acoustics) transforms in the same way in both approaches.

Finally, it is worth stressing here that the acoustic wave equations come from Newtonian hydrodynamics. In this conceptualization, the functions $p$ and $\phi_1$ are both scalar quantities under any change of coordinates. It is the Newtonian conceptualization versus the relativistic conceptualization of the wave equation in an abstract space what makes the respective transformations different. However, note that  in a relativistic scenario, when writing $p$ one would have to say to which object one is referring to, the trace of the spatial part of the stress-energy tensor or the contraction of the stress-energy tensor with the appropriate tetrads.

\section{\label{s:Conclusion}Conclusion}

In this paper we have deepened our understanding of analogue transformations applied to acoustics (ATA). Building on the results of Ref.~\onlinecite{GAR13-X}, we have clarified several fundamental differences between this technique and standard transformation acoustics (STA).
First, we have derived the set of transformations that do not preserve the form of the velocity potential equation in the original laboratory space. For these transformations, it is indispensable to apply the ATA method by constructing an auxiliary relativistic analogue spacetime before mapping back to the laboratory spacetime. As a second result, we have shown that the pressure wave equation commonly used in STA is not suitable for building an analogue transformational method, and have highlighted the importance of the background velocity. Third, we have examined in detail a spacetime transformation that can only be performed with ATA and that allows us to design acoustic frequency converters. As a fourth contribution, we have analyzed the case of spatial conformal transformations for which both approaches (ATA and STA) can be used, explaining why the mass density transforms differently in each method, whereas the speed of sound transforms equally. In this context, we have also obtained and compared the pressure transformation rules for both cases.

Overall, these results confirm the conclusion given in Ref.~\onlinecite{GAR13-X}: the different requirements that ATA imposes on acoustic metamaterial design compared to the standard approaches, and even more so: the potential of ATA to design applications which could so far not be obtained with any other methodology, open a completely new perspective on the field of Transformation Acoustics.

\begin{acknowledgments}
This work was developed under the framework of the ARIADNA contracts 4000104572/11/NL/KML and
4000104572/12/NL/KML of the European Space Agency. C.~G.-M., J.~S.-D., and A.~M. also acknowledge
support from Consolider EMET project (CSD2008-00066), A.~M. from project TEC2011-28664-C02-02,
and C.~B. and G.~J. from the project FIS2008-06078-C03-01.
\end{acknowledgments}

\appendix*

\section{General transformations for the acoustic equation} \label{s:Appendix}
One of the key issues in transformation methods is the range of coordinate transformations over which the relevant field equations are form invariant, see e.g.~\cite{Milton,NOR11-WM,Bergamin}. Here, we explicitly derive the conditions for coordinate transformations under which the acoustic wave equation preserves its form.

Take a specific acoustic equation for the velocity potential. For convenience let us write it in the form
\begin{equation}
\sqrt{-f} \partial_\mu f^{\mu\nu} \partial_\nu \phi=0~.
\label{acousticf}
\end{equation}
The coefficients $f^{\mu\nu}(t,\x)$ have a specific functional dependence in the Cartesian coordinates $(t,x)$. We have arranged them in the form of the inverse of a matrix array $f_{\mu\nu}$ and $f$ represents the determinant of this matrix of coefficients.

The equation does not incorporate by itself properties associated to changes of coordinates. This is true because without additional information we don't know the transformation properties of the coefficients. For instance, the $f^{\mu\nu}(t,\x)$ could be just an array of scalars, then a transformation of coordinates will involve only to take due care of the derivatives. The equation of acoustics we are dealing with comes from an initially Newtonian system. That is why we know that $f^{00}$ is a scalar, $f^{0i}$ a vector and $f^{ij}$ a tensor, all under spatial coordinate transformations. Time is an external independent parameter. Under changes of the time parameter all the coefficients should transform as scalars. Recall also that the field $\phi$ is a scalar under any change of coordinates.

Let us perform a general transformation of the acoustic equation to see its new form. Consider the form
\begin{equation}
\partial_t f^{00} \partial_t \phi + \partial_t f^{0i} \partial_i \phi + \partial_i f^{i0} \partial_t \phi + \partial_i f^{ij} \partial_j \phi=0~,
\end{equation}
or changing notation (renaming the label $(t,\x)$ by $(\bar{t},\bar{\x})$),
\begin{equation}
-\partial_{\bar{t}} \Phi \partial_{\bar{t}} \phi - \partial_{\bar{t}} V^{\bar{i}} \partial_{\bar{i}} \phi - \partial_{\bar{i}} V^{\bar{i}} \partial_{\bar{t}} \phi + \partial_{\bar{i}} f^{\bar{i} \bar{j}} \partial_{\bar{j}} \phi=0~.
\end{equation}
A change of coordinates (from $(\bar{t},\bar{\x})$ to $(t,\x)$) affects the derivatives in the following way
\begin{eqnarray}
&&\partial_{\bar{i}} = T_{\bar{i}}^i \partial_i + Z_{\bar{i}} \partial_t~;
\\
&&\partial_{\bar{t}} = W^i \partial_i + Z \partial_t~,
\end{eqnarray}
where
\begin{eqnarray}
&&T_{\bar i}^i := {\partial x^i \over \partial x^{\bar i}}~,~~~~
Z_{\bar i} :={\partial t \over \partial x^{\bar i}}~,
\\
&&W^i := {\partial x^i \over \partial {\bar t}}~,~~~~Z := {\partial t \over \partial \bar t}~.
\end{eqnarray}

Let us now proceed term by term with manipulations associated with the transformation of coordinates. We will signal with the symbols $\bar{i} \bar{j})$, $\bar{i} \bar{t})$, etc. the terms containing partial devivatives $\partial_{\bar i} \partial_{\bar j}$, $\partial_{\bar i} \partial_{\bar t}$, etc. respectively.
\begin{align} \nonumber
\bar i \bar j)~~~~~~\partial_{\bar i} f^{\bar i \bar j} \partial_{\bar j} \phi&=
Z_{\bar i} \partial_t f^{\bar i \bar j}  T_{\bar j}^j \partial_j \phi
+T_{\bar i}^i \partial_i f^{\bar i \bar j} Z_{\bar j} \partial_t \phi  \\
\label{ij2}
&+Z_{\bar i} \partial_t f^{\bar i \bar j} Z_{\bar j} \partial_t \phi
+ T_{\bar i}^i \partial_i f^{\bar i \bar j} T_{\bar j}^j \partial_j \phi~;
\end{align}
The last term in the previous expression can be rewritten as
\begin{equation}
T_{\bar i}^i \partial_i f^{\bar i \bar j} T_{\bar j}^j \partial_j \phi=
\nabla_i f^{ij} \nabla_j \phi= {1 \over \sqrt{\gamma}} \partial_i
\sqrt{\gamma} f^{ij} \partial_j \phi~,
\end{equation}
where we have introduced a spatial metric $\gamma_{ij}$, which is the Euclidean metric $\delta_{\bar i \bar j}$ written in arbitrary spatial coordinates.
\begin{align}
\bar t \bar i)~~~~~~\partial_{\bar t} V^{\bar i} \partial_{\bar i} \phi &=
W^i \partial_i V^{\bar i} T_{\bar i}^j \partial_j \phi
+ Z \partial_t V^{\bar i} T_{\bar i}^j \partial_j \phi\nonumber
\\
&
~~ +W^i \partial_i V^{\bar i} Z_{\bar i} \partial_t \phi
+ Z \partial_t V^{\bar i} Z_{\bar i} \partial_t \phi\nonumber\\
& =W^i \partial_i V^{j} \partial_j \phi
+ Z \partial_t V^{j} \partial_j \phi\nonumber \\
&
~~ +W^i \partial_i V^{\bar i} Z_{\bar i} \partial_t \phi
+ Z \partial_t V^{\bar i} Z_{\bar i} \partial_t \phi~.
\label{ti2}
\end{align}
\begin{align}
\bar i \bar t)~~~~~~\partial_{\bar i} V^{\bar i} \partial_{\bar t} \phi&=
T_{\bar i}^i \partial_i V^{\bar i} W^j \partial_j \phi
+ T_{\bar i}^i \partial_i V^{\bar i} Z \partial_t \phi
\nonumber \\
&
+Z_{\bar i} \partial_t V^{\bar i} W^j \partial_j \phi
+Z_{\bar i} \partial_t V^{\bar i} Z \partial_t \phi~.
\label{it2}
\end{align}
\begin{align}
\bar t \bar t)~~~~~~\partial_{\bar t} \Phi \partial_{\bar t} \phi&=
Z \partial_t \Phi Z \partial_t \phi
+ W^i \partial_i \Phi Z \partial_t \phi
\nonumber \\
&
+Z \partial_t \Phi W^j \partial_j \phi
+W^i \partial_i \Phi W^j \partial_j \phi~.
\label{tt2}
\end{align}
Consider now terms of the type $ij)$. From (\ref{ti2})
\begin{align}
&W^i \partial_i V^{j} \partial_j \phi = {1 \over \sqrt{\gamma}} \sqrt{\gamma} W^i \partial_i V^{j} \partial_j \phi
\nonumber
\\
&={1 \over \sqrt{\gamma}} \partial_i \sqrt{\gamma} W^i  V^{j} \partial_j \phi - {1 \over \sqrt{\gamma}} \partial_i (\sqrt{\gamma} W^i) V^{j} \partial_j \phi~.
\end{align}
If we the impose condition $\nabla_i W^i=0$, we obtain
\begin{eqnarray}
W^i \partial_i V^{j} \partial_j \phi  ={1 \over \sqrt{\gamma}} \partial_i \sqrt{\gamma} W^i  V^{j} \partial_j \phi~.
\end{eqnarray}
The same can be done with the last term in (\ref{tt2}):
\begin{eqnarray}
W^i \partial_i \Phi W^j \partial_j \phi={1 \over \sqrt{\gamma}} \partial_i \sqrt{\gamma} W^i  \Phi W^j \partial_j \phi~.
\label{ijp}
\end{eqnarray}
From the first term in (\ref{it2}) we have
\begin{eqnarray}
T_{\bar i}^i \partial_i V^{\bar i} W^j \partial_j \phi =
{1 \over \sqrt{\gamma}} \partial_i \sqrt{\gamma} V^i W^j  \partial_j \phi~.
\end{eqnarray}

There are several terms of the form $it)$. Considering again the condition $\nabla_i W^i=0$, we have from (\ref{tt2}) and (\ref{it2}) respectively:
\begin{eqnarray}
W^i \partial_i \Phi Z \partial_t \phi = {1 \over \sqrt{\gamma}}
\partial_i \sqrt{\gamma} W^i \Phi Z \partial_t \phi~;
\end{eqnarray}
\begin{eqnarray}
T_{\bar i}^i \partial_i V^{\bar i} Z \partial_t \phi=
{1 \over \sqrt{\gamma}}
\partial_i \sqrt{\gamma} V^{\bar i} Z \partial_t \phi~.
\end{eqnarray}
The terms of the form $ti)$ are
\begin{eqnarray}
Z \partial_t \Phi W^j \partial_j \phi~
\end{eqnarray}
from (\ref{tt2}), and
\begin{eqnarray}
Z \partial_t V^j \partial_j \phi~
\end{eqnarray}
from (\ref{ti2}).
If we now impose $\partial_t \sqrt{\gamma}=0$, we can rewrite them as
\begin{eqnarray}
{1 \over \sqrt{\gamma}} Z \partial_t \sqrt{\gamma} \Phi W^j \partial_j \phi~;
\end{eqnarray}
\begin{eqnarray}
{1 \over \sqrt{\gamma}} Z \partial_t \sqrt{\gamma} V^j \partial_j \phi~.
\end{eqnarray}
Notice that if we alternatively impose $\partial_i \sqrt{\gamma}=0$, then all the $\sqrt{\gamma}$ terms in the previous equations disappear, so that we do not need to additionally impose $\partial_t \sqrt{\gamma}=0$
to recover the initial acoustic form.

If we impose now condition $\partial_t Z_{\bar i}=0$, we can rewrite the corresponding terms in
\eqref{ij2} as
\begin{align}
&Z_{\bar i} \partial_t f^{\bar i \bar j}  T_{\bar j}^j \partial_j \phi
+Z_{\bar i} \partial_t f^{\bar i \bar j} Z_{\bar j} \partial_t \phi
\nonumber \\
&=\partial_t Z_{i} f^{ij} \partial_j \phi
+\partial_t Z_{i} f^{ij} Z_{j} \partial_t \phi
\nonumber \\
&={1 \over \sqrt{\gamma}}\partial_t \sqrt{\gamma} Z_{i} f^{ij} \partial_j \phi
+{1 \over \sqrt{\gamma}}\partial_t \sqrt{\gamma} Z_{i} f^{ij} Z_{j} \partial_t \phi~.
\end{align}
If we also impose the condition $\partial_i Z=0$, then we can rewrite (\ref{ijp})
\begin{equation}
{1 \over \sqrt{\gamma}} \partial_i \sqrt{\gamma} W^i  V^{j} \partial_j \phi=
{Z \over \sqrt{\gamma}} \partial_i Z^{-1} \sqrt{\gamma} W^i  V^{j} \partial_j \phi~,
\end{equation}
and equivalently other similar terms
\begin{equation}
{1 \over \sqrt{\gamma}} \partial_i \sqrt{\gamma} W^i \Phi Z \partial_t \phi=
{Z \over \sqrt{\gamma}} \partial_i \sqrt{\gamma} W^i \Phi \partial_t \phi~;
\end{equation}
\begin{equation}
T_{\bar i}^i \partial_i f^{\bar i \bar j} Z_{\bar j} \partial_t \phi=
{Z \over \sqrt{\gamma}}\partial_i \sqrt{\gamma} Z^{-1} f^{ij} Z_j \partial_t \phi~;
\end{equation}
\begin{equation}
W^i \partial_i V^{\bar i} Z_{\bar i} \partial_t \phi=
{Z \over \sqrt{\gamma}} \partial_i \sqrt{\gamma} Z^{-1} W^i V^{i} Z_{i} \partial_t \phi~.
\end{equation}
To be able to rewrite the terms
\begin{eqnarray}
{1 \over \sqrt{\gamma}}\partial_t \sqrt{\gamma} Z_{i} f^{ij} \partial_j \phi
+{1 \over \sqrt{\gamma}}\partial_t \sqrt{\gamma} Z_{i} f^{ij} Z_{j} \partial_t \phi~,
\end{eqnarray}
from \eqref{ij2} as
\begin{equation}
{Z \over \sqrt{\gamma}}\partial_t \sqrt{\gamma} Z^{-1} Z_{i} f^{ij} \partial_j \phi
+{Z \over \sqrt{\gamma}}\partial_t \sqrt{\gamma} Z^{-1} Z_{i} f^{ij} Z_{j} \partial_t \phi~,
\end{equation}
one needs $\partial_t Z=0$. However, notice that these terms will not exist if $Z_i=0$.

Finally, looking at all these conditions, we can conclude that to maintain the form of the acoustic equation, we need either of the following sets of conditions:
\begin{eqnarray}
&&W_i=0, ~~ Z_{\bar i}=0; \label{Constraint_1} \\
&&W_i \neq 0, ~~ Z_{\bar i}=0, \nabla_i W^i=0,~~\partial_t Z= 0 \mbox{ or } \partial_i Z= 0 ,\nonumber \\
&&\partial_t \sqrt{\gamma} =0  \mbox{ or } \partial_i\sqrt{\gamma}=0; \label{Constraint_2} \\
&&W_i= 0, ~~ Z_{\bar i} \neq 0, ~~ \partial_t Z_{\bar i}=0,~~ \partial_t Z= 0, \nonumber \\
&&\partial_t \sqrt{\gamma} =0  \mbox{ or } \partial_i\sqrt{\gamma}=0; \label{Constraint_3}\\
&&W_i \neq 0, ~~ Z_{\bar i} \neq 0, ~~ \nabla_i W^i=0,~~ \partial_t Z_{\bar i}= 0,~~ \partial_t Z= 0, \nonumber \\
&&\partial_t \sqrt{\gamma} =0  \mbox{ or } \partial_i\sqrt{\gamma}=0. \label{Constraint_4}
\end{eqnarray}

For example, from $Z_{\bar i}=0$ we directly obtain transformations of the form
\begin{eqnarray}
t = f({\bar t})~.
\end{eqnarray}
Alternatively, with $Z_{\bar i} \neq 0$, $\partial_t Z=0=\partial_i Z$, we obtain transformations of the form
\begin{eqnarray}
t = C{\bar t} + f({\bar \x})~.
\end{eqnarray}
In both cases, the space transformations
\begin{eqnarray}
\x = f({\bar t},{\bar \x}),
\end{eqnarray}
have to be such that $\nabla_i W^i=0$, $\partial_t \sqrt{\gamma}=0$ or $\partial_i \sqrt{\gamma}=0$. We can check that a Galilean transformation
\begin{eqnarray}
t=f(\bar{t})~;~~~~\x = \bar{\x} + \v \bar{t}~
\end{eqnarray}
satisfies the conditions. However, the frequency converter transformation in the main part of the article does not satisfy $\partial_i Z =0$. A contracting transformation of the form $\x=f(\bar{t}) \bar{\x}$ does not satisfy the condition
$\nabla_i W^i=0$.

In the case with more free parameters (\ref{Constraint_4}) the new equation can be written as
\begin{equation}
{Z \over \sqrt{\gamma}}\left(
-\partial_t {\tilde \Phi} \partial_{t} \phi - \partial_{t} {\tilde V}^{i} \partial_{i} \phi - \partial_{i} {\tilde V}^{i} \partial_{t} \phi + \partial_{i} {\tilde f}^{ij} \partial_{j} \phi \right) =0~,
\label{acoustict}
\end{equation}
with
\begin{equation}
{\tilde \Phi} = \sqrt{\gamma} \left( \Phi Z - Z_i f^{ij} Z_j Z^{-1}+ 2 V^i Z_i \right) ~;
\end{equation}
\begin{equation}
{\tilde V}^i = \sqrt{\gamma} \left(V^i + W^i \Phi - f^{ij} Z_j Z^{-1} + W^i V^j Z_j Z^{-1}\right)~;
\end{equation}
\begin{equation}
{\tilde f}^{ij} = \sqrt{\gamma} Z^{-1} \left(f^{ij} - W^i V^j -  V^i W^j  - \Phi W^i W^j \right)~.
\end{equation}
Multipliying by an appropriate constant, one will be able to write the tranformed equation (\ref{acoustict}) in the initial acoustic form (\ref{acousticf}).

\end{document}